\documentclass[floatfix, twocolumn, 10pt, aps, prd, showpacs, amsmath, amssymb, superscriptaddress]{revtex4-2}
\usepackage{graphicx}
\usepackage{hyperref}
\usepackage{subeqnarray}
\usepackage{color}
\usepackage[normalem]{ulem}
\usepackage{url}
\usepackage{multirow}
\usepackage{array}
\usepackage{subfigure}
\hypersetup{colorlinks=false}
\hypersetup{linktocpage}
\definecolor{darkblue}{RGB}{16,78,139}
\definecolor{darkgreen}{rgb}{0.04, 0.7, 0.2}
\newcommand{\beq}{\begin{equation}}
\newcommand{\eeq}{\end{equation}}
\newcommand{\beqn}{\begin{eqnarray}}
\newcommand{\eeqn}{\end{eqnarray}}
\newcommand{\beqs}{\begin{subeqnarray}}
\newcommand{\eeqs}{\end{subeqnarray}}
\newcommand{\nn}{\nonumber}

\begin{document}
\title{Ergoregion instability in a fluid with vorticity}

\author{Leandro A. Oliveira}
\email{laoliveira@ufpa.br}
\affiliation{Campus Universit\'ario Salin\'opolis, Universidade
Federal do Par\'a, 68721-000, Salin\'opolis, Par\'a, Brazil.}

\author{Carolina L. Benone}
\email{benone@ufpa.br}
\affiliation{Campus Universit\'ario Salin\'opolis, Universidade
Federal do Par\'a, 68721-000, Salin\'opolis, Par\'a, Brazil.}

\author{Lu\'is C. B. Crispino}
\email{crispino@ufpa.br}
\affiliation{Faculdade de F\'isica, Universidade
Federal do Par\'a, 66075-110, Bel\'em, Par\'a, Brazil.}

\date{\today}

\begin{abstract}
We investigate perturbations in a rotational and incompressible fluid flow. Interested in the phenomenon analogous to the black hole ergoregion instability, we verify the influence of the vorticity in the instability associated with this fluid system, in the presence of a region in which the fluid flow velocity is greater than the speed of the perturbation. With this aim, we compute the quasinormal modes of the system, using two different numerical methods, obtaining an excellent numerical agreement between them. We find that the vorticity tends to diminish the ergoregion instability of the system.

\end{abstract}


\maketitle

\section{Introduction}
Black holes (BHs) have come to play a central role is astrophysics, being related to the formation of galaxies~\cite{Cattaneo:2009ub} and presenting a possible ground to look for quantum structure signatures~\cite{Giddings:2017jts}. The first indirect observational evidences of BHs date back to the 1960s, with the discovery of quasars~\cite{Schmidt:1963wkp}. Recently, the first detection of gravitational wave was realized by the LIGO collaboration~\cite{LIGOScientific:2016aoc}. Up to the end of the third run of observations, 90 signals of gravitational waves had been detected~\cite{LIGOScientific:2021djp}. These detections are consistent with binary BHs, neutron star-black hole binaries and binary neutron stars.

The signal of gravitational waves can be divided in three parts: Inspiral, merger and ringdown. The ringdown phase is dominated by the quasinormal (QN) modes, which are characteristic modes of vibration. Due to the presence of the event horizon, these perturbations are, in general, naturally decaying, as the modes leak into the black hole through the event horizon.

The first investigations on QN modes (QNMs) of BHs were made by Vishveshwara, who studied the scattering of wave packets by a Schwarzschild black hole~\cite{Vishveshwara:1970zz}. Later, it was understood that most BH dynamical processes excite these modes (see, e.g., Ref.~\cite{Kokkotas:1999bd} and references therein). Moreover, as the frequencies of these modes depend only on the parameters of the BH, they act as an imprint of this object.

Since they arise in dynamical processes, QNMs are important to investigate the stability of BHs~\cite{Regge:1957td, Vishveshwara:1970cc, Zerilli:1970wzz, Detweiler:1973zz, Friedman:1974nqm, Press:1973zz, Detweiler:1980gk, Zouros:1979iw, Detweiler:1980uk}. Furthermore, the QN frequencies (QNFs) depend only on the parameters of the BH, such that they can be used as a test for the no-hair theorem~\cite{Dreyer:2003bv,Meidam:2014jpa}. QNMs may be used to investigate other systems, such as analogue acoustic holes~\cite{Berti:2004ju, Cardoso:2004fi, Dolan:2010zza, Dolan:2011ti, Oliveira:2015vqa, OC:2018, Barcelo:2007ru}.

An analogue event horizon arises, in analogue acoustic holes, when the radial background velocity of the fluid becomes equal to the speed of sound, i.e., when the Mach's number is equal to the unity~\cite{Unruh:1980cg, Visser:1997ux, AMproc, Barcelo:2005fc, Visser:2001fe}. On such systems, a perturbation in the fluid perceives and effective curved geometry, following a curved trajectory~\cite{Fischer:2001jz}. For surface waves, one can also find an event horizon by considering the radial background velocity equal to the speed of the perturbation~\cite{Schutzhold:2002rf,Volovik:2006}

The existence of an analogue event horizon gives rise to a diversity of phenomena, such as the absorption~\cite{Crispino:2007zz, Oliveira:2010zzb} and scattering of sound waves~\cite{Dolan:2009zza}, an analogous to the Aharonov-Bohm effect in a rotating fluid~\cite{Dolan:2011zza} and acoustic clouds around a black hole analogue in fluids~\cite{Benone:2014nla}.

The existence of an ergoregion both in spacetimes of General Relativity and acoustic analogues leads to the occurrence of events associated with the rotation of the system. Among them, we may point out the ergoregion instability, that is associated with unstable modes in a system with an ergoregion but without an event horizon~\cite{Friedman:1978, Comins:1978, Cardoso:2007az, Chirenti:2008pf}. In analogue models in fluids, this phenomenon has been investigated considering purely rotating systems~\cite{Oliveira:2014oja, Oliveira:2016adj, Oliveira:2018ckz, Hod:2014hda}.

Analogue models in fluids, first described by William Unruh in 1981, were initially obtained by assuming irrotationality in fluid flow~\cite{Unruh:1980cg}. We can also describe analogue models considering nonvanishing vorticity by assuming a non-Riemannian effective spacetime~\cite{GarciadeAndrade:2004au, GarciadeAndrade:2005ap}. Analogue BHs in fluids with vorticity have been recently described in~\cite{Patrick:2020qxd, PerezBergliaffa:2001nd}. An investigation of QNMs in a draining bathtub vortex with vorticity was considered in~\cite{Patrick:2018orp}. 

In this work we consider perturbations in a vortex fluid flow with vorticity. We compute the QNMs for this system using the direct integration method and the continued fraction method.
The remainder of this paper is structured as follows. In Sec.~\ref{sec-Perturbations} we study the propagation of linear perturbations in a rotational and incompressible fluid flow, using the description in the frequency domain. In Sec.~\ref{sec-Quasinormal} we describe the methods that we use to obtain the QNMs of this system, namely, direct integration (DI) and continued-fraction (CF) methods. In Sec.~\ref{sec-Results} we obtain the QNMs for this rotational system, validating and commenting our results, comparing the QNM frequencies obtained via DI and CF methods. We conclude with a brief discussion in Sec.~\ref{sec-Conclusion}.

\section{Perturbations in a fluid with vorticity}
\label{sec-Perturbations}
The non-perturbed fluid is represented by an incompressible fluid flow
\beq
\nabla\cdot \vec{v}_0=0,
\eeq
satisfying the Navier-Stokes equation
\beq
\dfrac{D\vec{v}_0}{Dt}+\frac{1}{\rho_0}\nabla p_0-\nu_{\rm kv} \nabla^2  \vec{v}_0=0,
\eeq
where $\nu_{\rm kv}$ is the kinematic viscosity, $\vec{v}_0$ is the background velocity, $\rho_0$ is the background mass density, $p_0$ is the background pressure and the material derivative is given by $$\dfrac{D}{Dt}\equiv\dfrac{\partial}{\partial t}+\left(\vec{v}_0\cdot \nabla\right)\,.$$

Furthermore, the background fluid flow is assumed to be rotational, with
\beq
\vec{\omega}_0=\nabla \times \vec{v}_0,
\eeq
being the background vorticity.

The linear perturbations may be represented by the following equations
\beq
\frac{D\rho_1}{Dt}+\rho_0\nabla\cdot\vec{v}_1=0
\label{pert_cont}
\eeq
and 
\beq
\frac{D\vec{v}_1}{Dt}+\left(\vec{v}_1\cdot \nabla\right)\vec{v}_0+\frac{1}{\rho_0}\nabla p_1-\nu_{\rm kv} \nabla^2  \vec{v}_1=0,
\label{pert_navier}
\eeq
where the subscript~``$1$'' indicates the increment on the flow quantities due to the perturbations.

Assuming an isentropic fluid flow, we may write 
\beq
\rho_1=\frac{1}{c_{\rm s}^2}p_1,\nn
\eeq
where $c_{\rm s}$ is the speed of sound.

The existence of vorticity on the perturbations may be denoted from the vector increment of velocity~$\vec{v}_1$, using the Helmholtz-Hodge decomposition~\cite{PerezBergliaffa:2001nd}
\beq
\vec{v}_1=\nabla \Phi+\vec{\Omega}_1,
\label{helm}
\eeq
where~$\vec{\Omega}_1$ is a vector field with non-vanishing curl, i.e., $$\nabla \times \vec{\Omega}_1\neq0.$$ 

Considering that the flow is restricted to the plane~$(r,\theta)$, the background vorticity may be written as
\beq
\vec{\omega}_0=\omega_0\hat{k}\nn,
\eeq
and the vector field~$\vec{\Omega}_1$ as
\beq
\vec{\Omega}_1=\hat{k}\times\nabla\Psi=\tilde{\nabla}\Psi,
\label{cograd}
\eeq
where~$\omega_0$ is the modulus of the background vorticity and~$\tilde{\nabla}$ is the co-gradient operator~\cite{Patrick:2018orp}.

We shall consider only the scalar perturbation~$\Phi$, i.e., the increment on the vorticity due to the perturbation is neglected, but the background vorticity is non-zero. This regime applies when the~$\nabla \times \vec{\Omega}_1$ is much smaller than the frequency of the perturbation.

Substituting Eqs.~\eqref{helm} and~\eqref{cograd} into Eqs.~\eqref{pert_cont} and~\eqref{pert_navier}, we may write 
\beq
\left[\frac{1}{c_{\rm s}^2}\frac{D^2}{Dt^2}-\nabla^2\left(1+\frac{\nu_{\rm kv}}{c_{\rm s}^2}\frac{D}{Dt}\right)+\frac{\omega_0^2}{c_{\rm s}^2}\right]\Phi=0,
\label{pert_equa_1}
\eeq
as the equation governing the scalar perturbation $\Phi$~\cite{Patrick:2018orp, Patrick:2020qxd}.

Furthermore, we may assume that the kinematic viscosity is much smaller than the square speed of the sound, so that we may rewrite Eq.~\eqref{pert_equa_1} as
\beq
\left(\square+\frac{\omega_0^2}{c_{\rm s}^2}\right)\Phi=0,
\label{pert_equa_2}
\eeq
where $$\square\equiv \frac{1}{c_{\rm s}^2}\frac{D^2}{Dt^2}-\nabla^2.$$

We write the background velocity, in cylindrical coordinates, as 
\beq
\vec{v}_0=v_{\theta}(r)\hat{\theta},
\label{back_velo}
\eeq
where~$v_{\theta}$ is the azimuthal component of the background velocity, describing a rotational fluid flow. Then, the modulus of the background vorticity may be obtained from
\beq
\omega_0=\frac{1}{r}\frac{d}{dr}\left(rv_\theta\right).
\eeq

We shall consider an empirical expression, proposed by Rosenhead, for the background tangential velocity~\cite{Rosenhead, Hite, Patrick:2018orp, Vatistas, Mih}, namely:
\beq
v_\theta(r)=\frac{Cr}{r_0^2+r^2},
\label{ang_velo}
\eeq
where~$C$ is a constant, related to the circulation~$\Gamma(r)$, which can be obtained from~$C=\dfrac{1}{2\pi}\Gamma(r\to \infty)$, and~$r_0$ is the radius of the vortex core.

The modulus of the vorticity, associated to Eq.~\eqref{ang_velo}, is given by
\beq
\omega_0(r)=\frac{2Cr_0^2}{\left(r_0^2+r^2\right)^2}.
\label{backvort}
\eeq

There is a certain position at~$r=r_{\rm e}$, which is the radius where the flow velocity becomes equal to the speed of sound, i.e. the Mach number,
\beq
M\equiv \dfrac{|\vec{v}|}{c_{\rm s}},
\label{Mach_number}
\eeq
is equal to unity. The position~$r=r_{\rm e}$ may be located from
\beq
M=\left.\frac{|\vec{v}|}{c_{\rm s}}\right|_{r=r_{\rm e}}=1.
\label{ergo_cond}
\eeq
Then, we find the position~$r_{\rm e}$ to be
\beq
r_{\rm e\,\pm}=\frac{r_0}{2}\left(\alpha_{\rm circ}\pm\sqrt{\alpha_{\rm circ}^2-4}\right),
\label{ergo_radius}
\eeq
where 
\beq
\alpha_{\rm circ} \equiv \pm\dfrac{|C|}{c_{\rm s}r_0}
\label{adim}
\eeq
is a dimensionless circulation parameter.

Note that, from Eq.~\eqref{ergo_radius}, we may conclude that there is a limit to the existence of the position~$r_{\rm e\,\pm}$, which is~$|\alpha_{\rm circ}| \geq 2$. For~$|\alpha_{\rm circ}| = 2$ there is only one point in which~$M=1$, satisfying~$r_{\rm e\,+}=r_{\rm e\,-}=r_0$, while for~$|\alpha_{\rm circ}| > 2$ there are two points~($r_{\rm e\,-}$ and~$r_{\rm e\,+}$) in which~$M=1$, satisfying the following inequality~$r_{\rm e\,-} < r_0 < r_{\rm e\,+}$, i.e., the radius of the vortex core is located between the inner and outer boundary of the analogue ergoregion. Furthermore, note that the positive (negative) sign in Eq.~\eqref{adim} corresponds to counter-clockwise (clockwise) rotation. Henceforth, we set the rotation as counter-clockwise, i.e.,~$\alpha_{\rm circ} \geq 2$~\cite{Oliveira:2018ckz}.

From Eq.~\eqref{adim}, we may rewrite Eq.~\eqref{backvort}, using a dimensionless radial coordinate~$x=r/r_0$, a dimensionless background vorticity~$\varpi_0=\omega_0 r_0/c_{\rm s}$ and a dimensionless position~$x_{\rm e\pm} \equiv r_{\rm e\pm}/r_0 $, namely
\beq
\varpi_0=\frac{2\alpha_{\rm circ}}{\left(1+x^2\right)^2}.
\label{dimen_vort}
\eeq
Note that~$\varpi_0(x)$ goes to zero when~$x \to \infty$.

\begin{figure}[htpb!]
\centering
\includegraphics[width=0.5\textwidth]{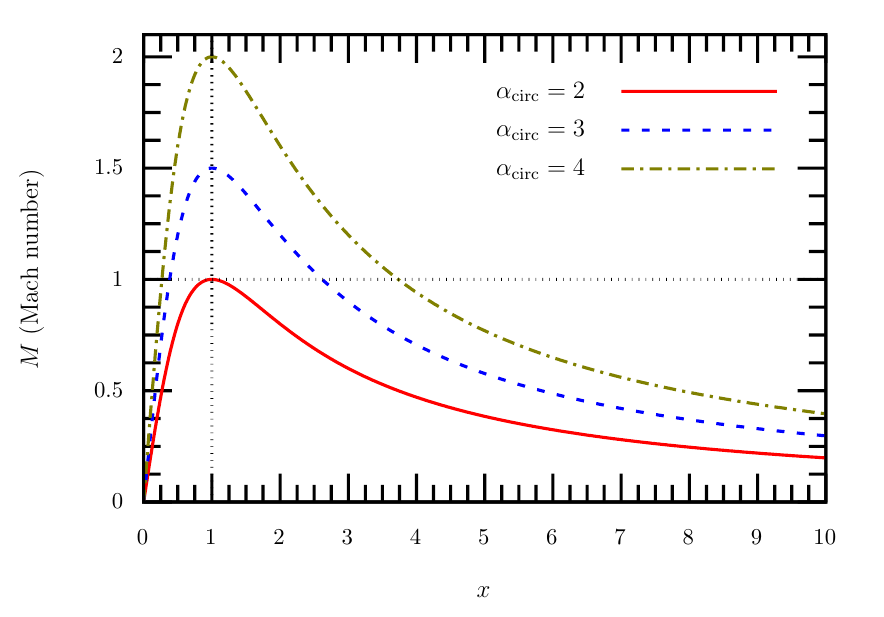}
\caption{The Mach number~$M$, as a function of the  dimensionless radial coordinate~$x$, for~$\alpha_{\rm circ}=2,\,3\,{\rm and}\,4$.}
\label{Mach}
\end{figure}

In Fig.~\ref{Mach} we plot the Mach number~$M$, defined in Eq.~\eqref{Mach_number}, as a function of the dimensionless radial coordinate~$x$, for different values of the dimensionless circulation parameter~$\alpha$. At the radius of the vortex core the background tangential velocity is maximal.

\begin{figure}[htpb!]
\centering
\includegraphics[width=0.5\textwidth]{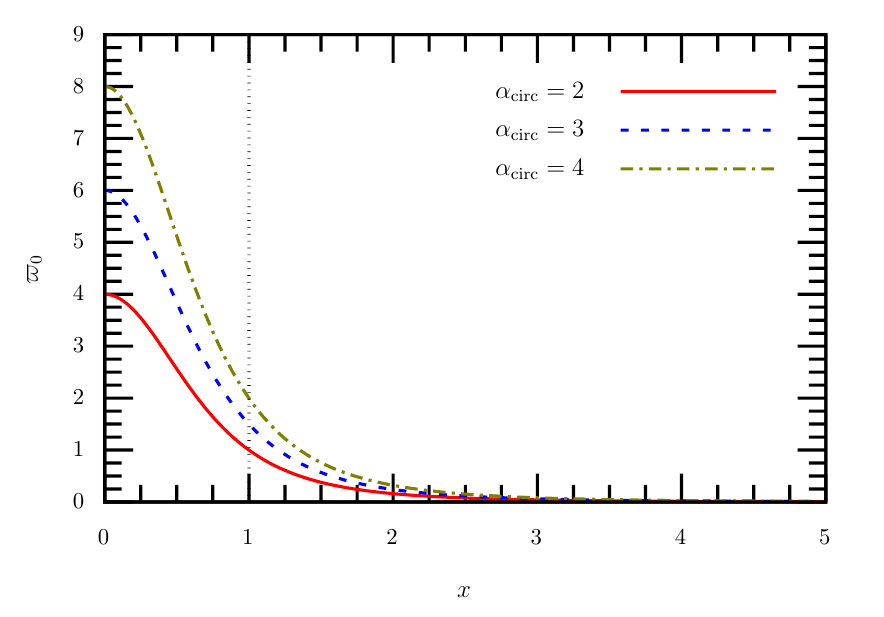}
\caption{Dimensionless background vorticity~$\varpi_0$, as a function of the  dimensionless radial coordinate~$x$, for~$\alpha_{\rm circ}=2,\,3\,{\rm and}\,4$.}
\label{dvort}
\end{figure}

In Fig.~\ref{dvort} we plot the dimensionless background vorticity~$\varpi_0$, as a function of the dimensionless radial coordinate~$x$, obtained from Eq.~\eqref{dimen_vort}.

Using the following {\it ansatz}
\beq
\Phi(t,r,\theta)=\frac{1}{\sqrt{r}}\sum_{m=-\infty}^{\infty}\phi_{\omega m}(r)\exp\left[i\left(m\theta-\omega t\right)\right],
\label{ansatz1}
\eeq
we may decompose the field~$\Phi$ in terms of azimuthal modes to exhibit the dependence on the frequency of the perturbations, obtaining the following radial wave equation
\beq
\left[\frac{d^2}{d r^2}+\frac{1}{c_{\rm s}^2}\left(\omega-\frac{mv_\theta}{r}\right)^2 -V_{\omega m}(r)\right]\phi_{\omega m}(r)=0,
\label{radial_wave}
\eeq
where~$m$ is an integer to require~$\theta$-periodicity,~$\omega$ is the frequency of the perturbations and the effective potential~$V_{m}(r)$ is
\beqn
V_{m}(r)=\frac{m^2-1/4}{r^2}+\frac{\omega_0^2}{c_{\rm s}^2}.\nn
\eeqn

We may rewrite Eq.~\eqref{radial_wave}, using a dimensionless frequency~$\varpi=\omega r_0/c_{\rm s}$ and Eq.~\eqref{adim}, namely
\beq
\left[\frac{d^2}{d x^2}+\left(\varpi-\frac{{\cal M} m}{x}\right)^2-\widetilde{V}_{m}(x)\right]\phi_{\varpi m}(x)=0,
\label{radial_wave_x}
\eeq
where the dimensionless effective potential~$\widetilde{V}_{m}(x)$ is
$$\widetilde{V}_{m}(x)=\frac{m^2-1/4}{x^2}+\varpi_0^2$$
and 
$${\cal M} = \frac{\alpha_{\rm circ}\,x}{1+x^2}.$$

In Eq.~\eqref{radial_wave_x}, when Eq.~\eqref{ang_velo} is considered, we can see that there are symmetries associated to the frequency~$\varpi$, relating the co-rotating modes~($Cm>0$) and the counter-rotating ones~($Cm<0$), namely~$\varpi({Cm>0}) = -\varpi^*({Cm<0})$, where~``$\,{}^*\,$'' denotes complex conjugation. Henceforth, considering these symmetries, we may assume, without loss of generality, that~$m > 0$ and~$C > 0$. This is possible since we know that to each QNM frequency of a co-rotating mode there is a corresponding one of a counter-rotating mode with opposite real part and the same imaginary part~\cite{Oliveira:2018ckz}.

\section{Quasinormal modes}
\label{sec-Quasinormal}

\subsection{Boundary conditions}
To find the QNFs,~$\varpi$, we assume two boundary conditions for~$\phi_{\varpi m}(x)$ at~$x\to \infty$ and at~$x=x_{\rm in}$, namely
\beqn
\phi_{\varpi m}\left(x\to \infty \right) \sim \exp{\left( {i\varpi x}\right)}\, ,
\label{BCI}
\eeqn
and
\beqn
\left[\frac{d}{dx}\left( \frac{\phi_{\varpi m}(x)}{\sqrt{x}}\right) \right]_{x=x_{\rm in}}=0.
\label{BCII}
\eeqn
The boundary condition given by Eq.~\eqref{BCI} is in accordance with the asymptotic behavior of Eq.~\eqref{radial_wave} at infinity. The boundary condition given by Eq.~\eqref{BCII} is one of Neumann type and describes a cutt-off on the radial component of the velocity~$\vec{v}_1$ at an inner radius of the vortex~$x=x_{\rm in}$~\cite{Lax:1948}.

We chose to impose the boundary condition at~$x=x_{\rm in}$ rather than at~$x=0$, as the latter implies a divergence in the differential equation, due to the axial symmetry of the problem. This choice also helps us to analyze the system under consideration, as we can vary the value of~$x_{\rm in}$ and verify how it affects our results.

\subsection{Direct integration method} 
To employ the direct integration method in determining the QNM frequencies, first, we may consider the following boundary conditions on the wave function~$\phi_{\varpi m}\left(x \right)$, according to Eq.~\eqref{BCI}, namely
\beqn
\phi_{\varpi m}\left(x\to \infty \right) = \exp\left( {i\varpi x}\right) \sum_{j=0}^{j_{\rm max}}{\frac{a_j}{x^j}}\,,
\label{series_di}
\eeqn
where $a_j$ are coefficients which can be determined from Eq.~\eqref{radial_wave_x}. Second, we integrate inwards Eq.~\eqref{radial_wave_x}, in the range~$\infty > x \geq x_{\rm in}$.

At $x=x_{\rm in}$, we extract the QNM frequencies as roots of the boundary condition given by Eq.~\eqref{BCII} using a standard root-finding algorithm, such as Newton's method~\cite{Chandrasekhar:1975zza}.

The results of the QNM frequencies obtained via direct integration method are exhibited in Sec.~\ref{sec-Results}.

\subsection{Continued fraction method} 
\label{sec-Continued}
We use the following ansatz~\cite{Leaver:1985ax}
\beq
\phi_{\varpi m}(x) = \exp\left(i\varpi x\right) \sum_{n=0} a_n \left( 1 -\frac{x_{\rm in}}{x} \right)^{n}\,.   
\label{series_cf}
\eeq
Substituting Eq.~\eqref{series_cf} into Eq.~\eqref{radial_wave_x}, we may find the following recurrence relations
\begin{widetext}
\beqn \label{recu_1}
&&\alpha_1a_2+\beta_1a_1 +\gamma_1a_0= 0, \nn\\
&&\alpha_2a_3+\beta_2a_2+\gamma_2a_1 +\delta_2a_0 = 0,  \nn\\
&&\alpha_3a_4+\beta_3a_3+\gamma_3a_2 +\delta_3a_1+\varepsilon_3a_0 = 0,  \nn\\
&&\alpha_4a_5+\beta_4a_4+\gamma_4a_3 +\delta_4a_2+\varepsilon_4a_1+\zeta_4a_0 = 0,  \nn\\
&&\alpha_5a_6+\beta_5a_5+\gamma_5a_4 +\delta_5a_3+\varepsilon_5a_2+\zeta_5a_1 +\eta_5 a_0 = 0,  \\
&&\alpha_6a_7+\beta_6a_6+\gamma_6a_5 +\delta_6a_4+\varepsilon_6a_3+\zeta_6a_2 +\eta_6 a_1+\lambda_6a_0 = 0,  \nn\\
&&\alpha_7a_8+\beta_7a_7+\gamma_7a_6 +\delta_7a_5+\varepsilon_7a_4+\zeta_7a_3 +\eta_7 a_2+\lambda_7a_1+\mu_7a_0 = 0,  \nn\\
&&\alpha_8a_9+\beta_8a_8+\gamma_8a_7 +\delta_8a_6+\varepsilon_8a_5+\zeta_8a_4 +\eta_8 a_3+\lambda_8a_2+\mu_8a_1+ \nu_8 a_0 = 0,  \nn\\
&&\alpha_n a_{n+1}+\beta_na_{n}+\gamma_n a_{n-1}+\delta_{n}a_{n-2}+\epsilon_{n}a_{n-3}+\zeta_{n}a_{n-4}+\eta_{n}a_{n-5} +\lambda_{n} a_{n-6}+\mu_{n}a_{n-7}+\nu_{n}a_{n-8}+\xi_{n}a_{n-9}= 0, \quad \mbox{for} \,\, n \geq 9,\nn
\eeqn
where the recurrence coefficients are 
\beqn
\alpha_n &=& 4 n (1 + n) (1 + x_{\rm in}^2)^4,\nn\\
\beta_n &=& -8 n (1 + 
   x_{\rm in}^2)^3 ((-1 + n) (5 + x_{\rm in}^2) + (1 + x_{\rm in}^2) (1 - 
      i x_{\rm in} \varpi)),\nn\\
\gamma_n &=& (1 + x_{\rm in}^2)^4 + 4 (-1 + n)^2 (1 + x_{\rm in}^2)^2 (45 + 22 x_{\rm in}^2 + x_{\rm in}^4) - 
 16 x_{\rm in}^2 \alpha_{\rm circ}^2 - 
 4 m^2 (1 + x_{\rm in}^2)^2 (1 + x_{\rm in}^4 - x_{\rm in}^2 (-2 + \alpha_{\rm circ}^2)) \nn\\
 &-& 
 8 m x_{\rm in}^2 (1 + x_{\rm in}^2)^3 \alpha_{\rm circ} \varpi + 
 4 (-1 + n) (1 + x_{\rm in}^2)^2 (-27 - 2 x_{\rm in}^2 + x_{\rm in}^4 - 
    16 i x_{\rm in} (1 + x_{\rm in}^2) \varpi),\nn\\
\delta_{n} &=& 8 (-(1 + x_{\rm in}^2) (85 + 58 x_{\rm in}^2 + 5 x_{\rm in}^4 + 
      4 (-1 + n)^2 (5 + x_{\rm in}^2) (3 + 2 x_{\rm in}^2) - 
      12 (-1 + n) (12 + 9 x_{\rm in}^2 + x_{\rm in}^4)) \nn\\
      &+& 12 x_{\rm in}^2 \alpha_{\rm circ}^2 + 
   m^2 (4 (1 + x_{\rm in}^2)^3 - 
      x_{\rm in}^2 (3 + 4 x_{\rm in}^2 + x_{\rm in}^4) \alpha_{\rm circ}^2) + 
   4 i (-2 + n) x_{\rm in} (1 + x_{\rm in}^2)^2 (7 + x_{\rm in}^2) \varpi + 
   6 m (x_{\rm in} + x_{\rm in}^3)^2 \alpha_{\rm circ} \varpi),\nn\\
\varepsilon_{n} &=& 4 (931 + 42 (-21 + 5 (-1 + n)) (-1 + n) + 1135 x_{\rm in}^2 + 
   m^2 (-4 (1 + x_{\rm in}^2)^2 (7 + x_{\rm in}^2) + 
      x_{\rm in}^2 (15 + 12 x_{\rm in}^2 + x_{\rm in}^4) \alpha_{\rm circ}^2) \nn\\
      &-& 
   6 m x_{\rm in}^2 (5 + 6 x_{\rm in}^2 + x_{\rm in}^4) \alpha_{\rm circ} \varpi + 
   x_{\rm in} (280 (-5 + n) (-1 + n) x_{\rm in} + 309 x_{\rm in}^3 + 
      30 (-11 + 3 (-1 + n)) (-1 + n) x_{\rm in}^3 \nn\\
      &+& (3 - 
         2 (-1 + n))^2 x_{\rm in}^5 - 60 x_{\rm in} \alpha_{\rm circ}^2 - 
      16 i (-3 + n) (1 + x_{\rm in}^2) (7 + 3 x_{\rm in}^2) \varpi)),\nn\\
\zeta_{n} &=& 8 (-1141 - 126 (-7 + n) (-1 + n) - 934 x_{\rm in}^2 + 
   2 m^2 (14 - 5 x_{\rm in}^2 (-4 + \alpha_{\rm circ}^2) - 
      2 x_{\rm in}^4 (-3 + \alpha_{\rm circ}^2)) \nn\\
      &+& 
   4 m x_{\rm in}^2 (5 + 3 x_{\rm in}^2) \alpha_{\rm circ} \varpi + 
   x_{\rm in} (-129 x_{\rm in}^3 + 
      2 (-1 + n) x_{\rm in} (322 + 48 x_{\rm in}^2 - (-1 + n) (56 + 9 x_{\rm in}^2))\nn\\
      &+& 
      40 x_{\rm in} \alpha_{\rm circ}^2
      + 
      2 i (-4 + n) (35 + 30 x_{\rm in}^2 + 3 x_{\rm in}^4) \varpi)),\nn\\
\eta_{n} &=& 2 (6419 + 84 (-39 + 5 (-1 + n)) (-1 + n) + 3166 x_{\rm in}^2 + 
   2 m^2 (-70 + 15 x_{\rm in}^2 (-4 + \alpha_{\rm circ}^2) + 
      2 x_{\rm in}^4 (-3 + \alpha_{\rm circ}^2)) \nn\\
      &-& 
   12 m x_{\rm in}^2 (5 + x_{\rm in}^2) \alpha_{\rm circ} \varpi + 
   x_{\rm in} (4 (-1 + n) x_{\rm in} (-420 + 56 (-1 + n) + 3 (-8 + n) x_{\rm in}^2) + 
      3 x_{\rm in} (49 x_{\rm in}^2 - 40 \alpha_{\rm circ}^2) \nn\\
      &-& 
      32 i (-5 + n) (7 + 3 x_{\rm in}^2) \varpi)),\nn\\
\lambda_{n} &=& 8 (-1387 + 12 (48 - 5 (-1 + n)) (-1 + n) - 343 x_{\rm in}^2 + 
   m^2 (28 - 3 x_{\rm in}^2 (-4 + \alpha_{\rm circ}^2)) + 
   6 m x_{\rm in}^2 \alpha_{\rm circ} \varpi \nn\\
   &+& 
   4 x_{\rm in} ((37 - 4 (-1 + n)) (-1 + n) x_{\rm in} + 3 x_{\rm in} \alpha_{\rm circ}^2 + 
      i (-6 + n) (7 + x_{\rm in}^2) \varpi)),\nn\\
\mu_{n} &=& 4 (1465 + 9 (-57 + 5 (-1 + n)) (-1 + n) + 121 x_{\rm in}^2 + 
   m^2 (-28 + x_{\rm in}^2 (-4 + \alpha_{\rm circ}^2)) - 
   2 m x_{\rm in}^2 \alpha_{\rm circ} \varpi \nn\\
   &+& 
   4 x_{\rm in} ((-12 + n) (-1 + n) x_{\rm in} - x_{\rm in} \alpha_{\rm circ}^2 - 
      4 i (-7 + n) \varpi)),\nn\\
\nu_{n} &=& 8 (-218 + 4 m^2 + 66 (-1 + n) - 5 (-1 + n)^2 + 
   i (-8 + n) x_{\rm in} \varpi),\nn\\
\xi_{n} &=& -4 m^2 + (15 - 2 (-1 + n))^2.\nn
\eeqn
\end{widetext}

We apply eight Gaussian eliminations in Eq.~\eqref{recu_1} and obtain a three-term recurrence relation
\beqn
\alpha_n a_{n+1}+\beta_na_{n}+\gamma_n a_{n-1} = 0, \hspace{0.5cm} \mbox{for} \hspace{0.1cm} n \geq 1. 
\label{recu_2}
\eeqn
Manipulating this equation, we find that
\beq
\frac{a_1}{a_0} =-\dfrac{\gamma_1}{\beta_1-\dfrac{\alpha_1\gamma_2}{\beta_2-\dfrac{\alpha_2\gamma_3}{\beta_3-...}}}.
\label{a10rr}
\eeq
Considering Eq.~\eqref{series_cf}, we find that the boundary condition~\eqref{BCII} gives us
\beq
\frac{a_1}{a_0}= \frac{1}{2} - i \varpi x_{in}.
\label{a10bc}
\eeq
Combining Eqs.~\eqref{a10rr} and~\eqref{a10bc}, we find that the minimal solution is given by
\beqn
1 -2i\varpi x_{\rm in} +\dfrac{2\gamma_1}{\beta_1-\dfrac{\alpha_1\gamma_2}{\beta_2-\dfrac{\alpha_2\gamma_3}{\beta_3-...}}}=0.
\label{cont_frac}
\eeqn

\section{Results}
\label{sec-Results}
In Tables~\eqref{table1} and~\eqref{table2}, we exhibit some values of the QNM frequencies to test the accuracy of the direct integration and continued-fraction methods. We note that the methods are in excellent agreement for both stable and unstable cases.
\begin{table*}[hbtp!]
\centering \caption{QNM frequencies of the fundamental mode~($n=0$) for different azimuthal numbers~$m$ and different values of the dimensionless parameter~$\alpha_{\rm circ}$, for boundary condition position~$x_{\rm in}=0.75$. The number in the parenthesis corresponds to the precision.}
\vskip 10pt
\begin{tabular}{@{}ccccccc@{}}
\hline \hline
\multicolumn{5}{c}{$m=2$}\\ \hline\hline
\multicolumn{1}{c}{} & \multicolumn{2}{c}{$\alpha_{\rm circ}=2.0$}& \multicolumn{2}{c}{$\alpha_{\rm circ}=4.0$}\\ \hline
Method        &$\text{Re}(\varpi)$ &$\text{Im}(\varpi)$ & $\text{Re}(\varpi)$ & $\text{Im}(\varpi)$\\
DI                 &$ -0.2738630610(6)  $        & $ - 0.030470272(6) $      &$ +0.8519075782544  $  &$ +0.005030287188(6)  $     \\
CF                 &$ -0.2738630610(1) $          &$ - 0.030470272(5) $      &$ +0.8519075782544 $  &$ +0.005030287188(8) $  \\
\hline \hline
\multicolumn{5}{c}{$m=3$}\\ \hline\hline
\multicolumn{1}{c}{} & \multicolumn{2}{c}{$\alpha_{\rm circ}=2.0$}& \multicolumn{2}{c}{$\alpha_{\rm circ}=4.0$}\\ \hline
Method        &$\text{Re}(\varpi)$ &$\text{Im}(\varpi)$ & $\text{Re}(\varpi)$ & $\text{Im}(\varpi)$\\
DI                 &$ -0.313570188531(6) $          &$  - 0.003311230409(6) $       &$ +2.00568207201(1) $  &$ +0.002400344286(5) $     \\
CF                 &$ -0.313570188531(2)   $        &$  - 0.003311230409(3)  $      &$ +2.00568207201(0)  $  &$   +0.002400344286(4)$  \\
\hline \hline
\end{tabular}
\label{table1}
\end{table*}

\begin{table*}[hbtp!]
\centering \caption{QNM frequencies of the fundamental mode~($n=0$) for different azimuthal numbers~$m$ and different values of the dimensionless parameter~$\alpha_{\rm circ}$, for boundary condition position~$x_{\rm in}=1.0$. The number in the parenthesis corresponds to the precision.}
\vskip 10pt
\begin{tabular}{@{}ccccccc@{}}
\hline \hline
\multicolumn{5}{c}{$m=2$}\\ \hline\hline
\multicolumn{1}{c}{} & \multicolumn{2}{c}{$\alpha_{\rm circ}=2.0$}& \multicolumn{2}{c}{$\alpha_{\rm circ}=4.0$}\\ \hline
Method        &$\text{Re}(\varpi)$ &$\text{Im}(\varpi)$ & $\text{Re}(\varpi)$ & $\text{Im}(\varpi)$\\
DI                 &$ -0.2621556431(4)  $  & $ - 0.0277687428(5) $      &$ +0.748057341979(4)  $ &$ +0.0061307532280(4) $     \\
CF                 &$ -0.2621556431(5) $   & $ - 0.0277687428(0)  $     &$ +0.748057341979(5)   $  &$ +0.0061307532280(5) $  \\
\hline \hline
\multicolumn{5}{c}{$m=3$}\\ \hline\hline
\multicolumn{1}{c}{} & \multicolumn{2}{c}{$\alpha_{\rm circ}=2.0$}& \multicolumn{2}{c}{$\alpha_{\rm circ}=4.0$}\\ \hline
Method        &$\text{Re}(\varpi)$ &$\text{Im}(\varpi)$ & $\text{Re}(\varpi)$ & $\text{Im}(\varpi)$\\
DI                 &$  -0.319916092(4)  $          & $ - 0.00466066406(5) $       &$ +1.60652966308(3)  $  &$ +0.00291820963(9) $     \\
CF                 &$  -0.319916092(3)   $         & $ - 0.00466066406(7) $       &$ +1.60652966308(2)   $  &$ +0.00291820963(8)  $  \\
\hline \hline
\end{tabular}
\label{table2}
\end{table*}

\begin{figure*}[htpb!]
\centering
\subfigure[Real part of the fundamental~($n = 0$) QNM frequencies~$\varpi$, for~$\alpha_{\rm circ}=2.0$.]{%
\includegraphics[width=0.5\textwidth]{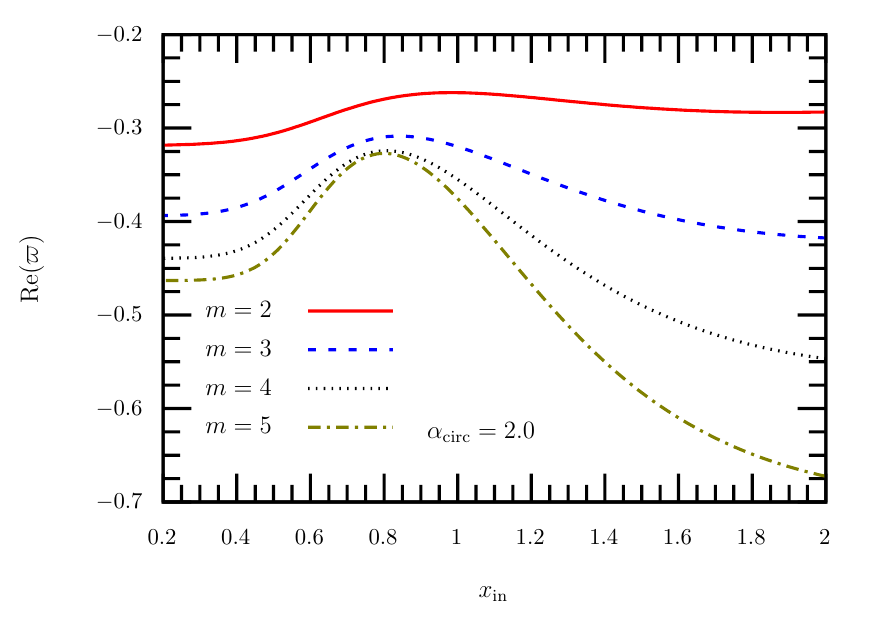}
\label{real_freq1}}\subfigure[Imaginary part of the fundamental~($n = 0$) QNM frequencies~$\varpi$, for~$\alpha_{\rm circ}=2.0$.]{%
\includegraphics[width=0.5\textwidth]{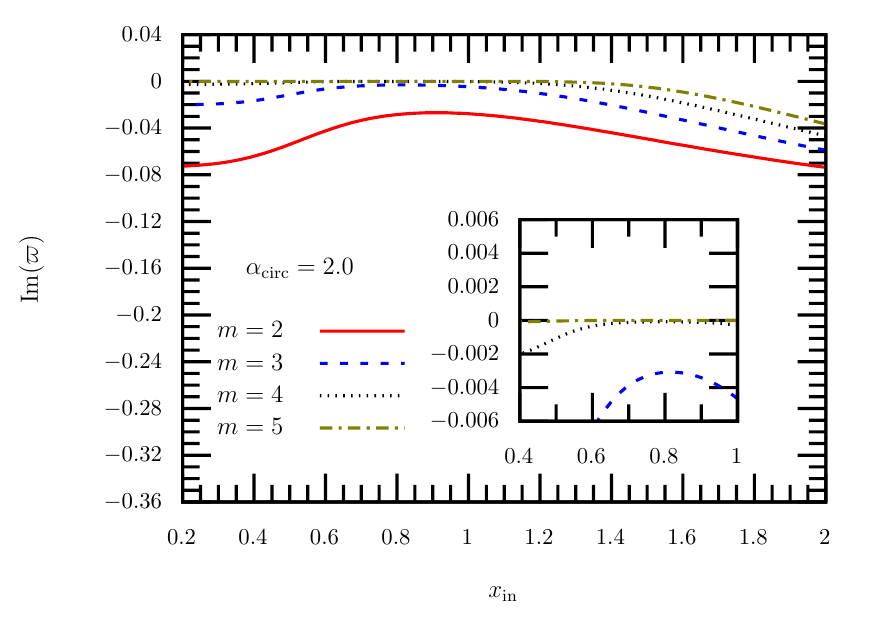}
\label{imag_freq1}}
\subfigure[Real part of the fundamental~($n = 0$) QNM frequencies~$\varpi$, for~$\alpha_{\rm circ}=4.0$.]{%
\includegraphics[width=0.5\textwidth]{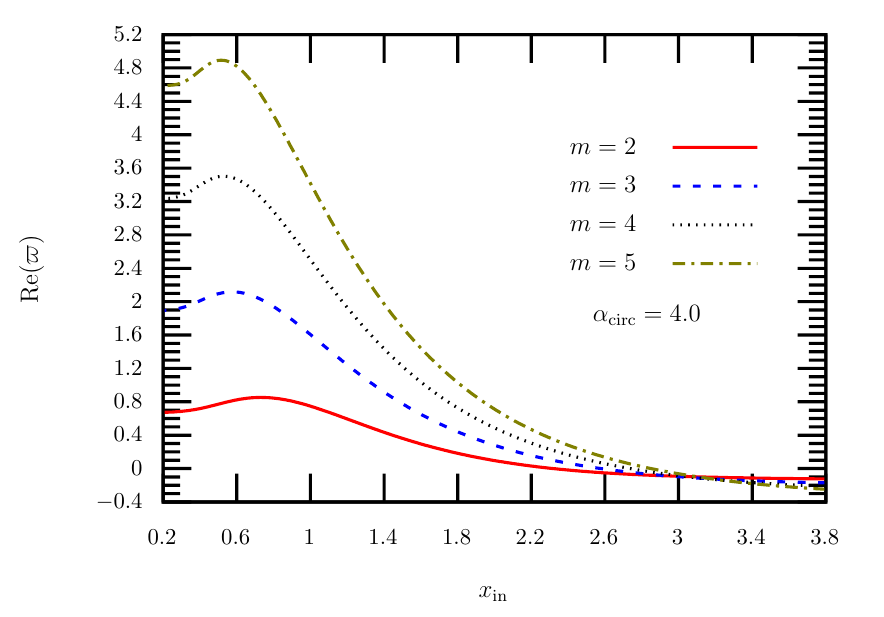}
\label{real_freq2}}\subfigure[Imaginary part of the fundamental~($n = 0$) QNM frequencies~$\varpi$, for~$\alpha_{\rm circ}=4.0$.]{%
\includegraphics[width=0.5\textwidth]{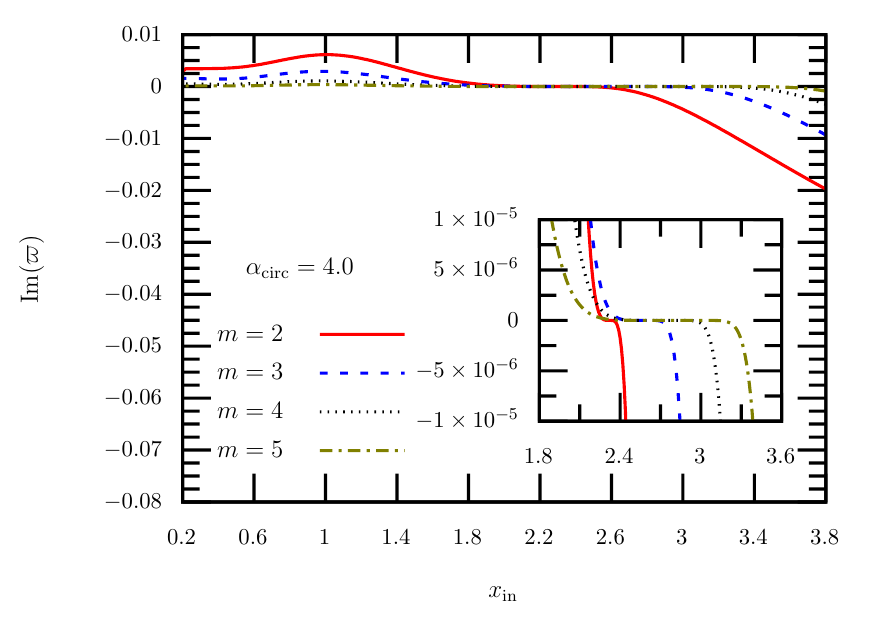}
\label{imag_freq2}}
\caption{Real and imaginary parts of the fundamental~($n = 0$) QNM frequencies~$\varpi$, for azimuthal numbers $m=\,2,\,3,\,4,\,5$ and~$\alpha_{\rm circ}=2.0\,\,{\rm and}\,\,4.0$, as functions of~$x_{\rm in}$, obtained via continued-fraction method.}
\label{Freqs1}
\end{figure*}

In Fig.~\ref{Freqs1} we plot the real (left plots) and imaginary (right plots) parts of the fundamental ($n = 0$) QNM frequencies~$\varpi$, as functions of~$x_{\rm in}$, for azimuthal numbers~$m=2,\,3,\,4,\,5,$ and $\alpha_{\rm circ}=2.0\,\,{\rm and}\,\,4.0$, obtained via continued-fraction method. For the extremal case there are only stable modes, as the ergoregion corresponds to a circle with radius~$r_{e+}=r_{e-}=r_0$.

\begin{figure*}[htpb!]
\centering
\subfigure[Real part of the fundamental~($n = 0$) QNM frequencies~$\varpi$, for~$x_{\rm in}=1.0$.]{%
\includegraphics[width=0.5\textwidth]{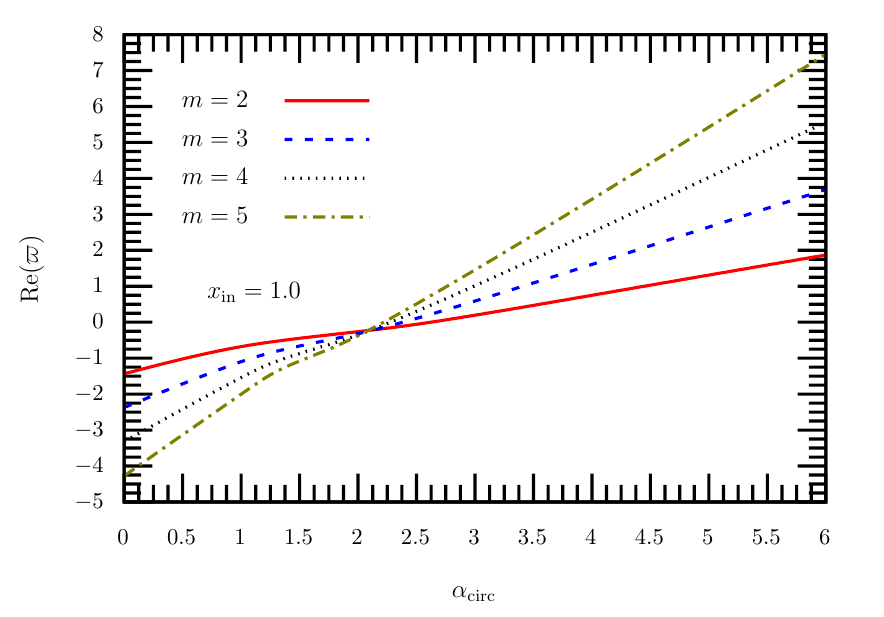}
\label{real_freq3}}\subfigure[Imaginary part of the fundamental~($n = 0$) QNM frequencies~$\varpi$, for~$x_{\rm in}=1.0$.]{%
\includegraphics[width=0.5\textwidth]{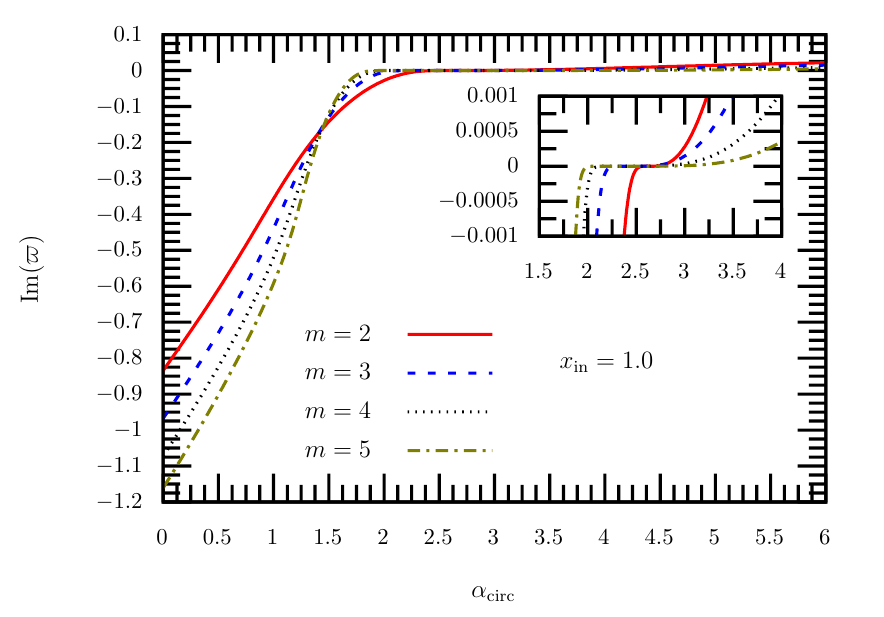}
\label{imag_freq3}}
\caption{Real and imaginary parts of the fundamental~($n = 0$) QNM frequencies~$\varpi$, for azimuthal numbers $m=\,2,\,3,\,4,\,5$ and $x_{\rm in}=1.0$, as functions of~$\alpha_{\rm circ}$, obtained via continued-fraction method.}
\label{Freqs2}
\end{figure*}

In Fig.~\ref{Freqs2} we plot the real (left plots) and imaginary (right plots) parts of the fundamental~($n = 0$) QNM frequencies~$\varpi$, as functions of~$\alpha_{\rm circ}$, for azimuthal numbers~$m=2,\,3,\,4,\,5,$ and~$x_{\rm in}=1.0$, obtained via continued-fraction method. Note that as the azimuthal number~$m$ is increased the threshold between stability and instability increases, denoting the dependence on large-$m$ values of the ergoregion instability phenomenon~\cite{Oliveira:2014oja, Oliveira:2016adj, Oliveira:2018ckz}.

\begin{figure*}[htpb!]
\centering
\subfigure[Real part of the fundamental~($n = 0$) QNM frequencies~$\omega$]{%
\includegraphics[width=0.5\textwidth]{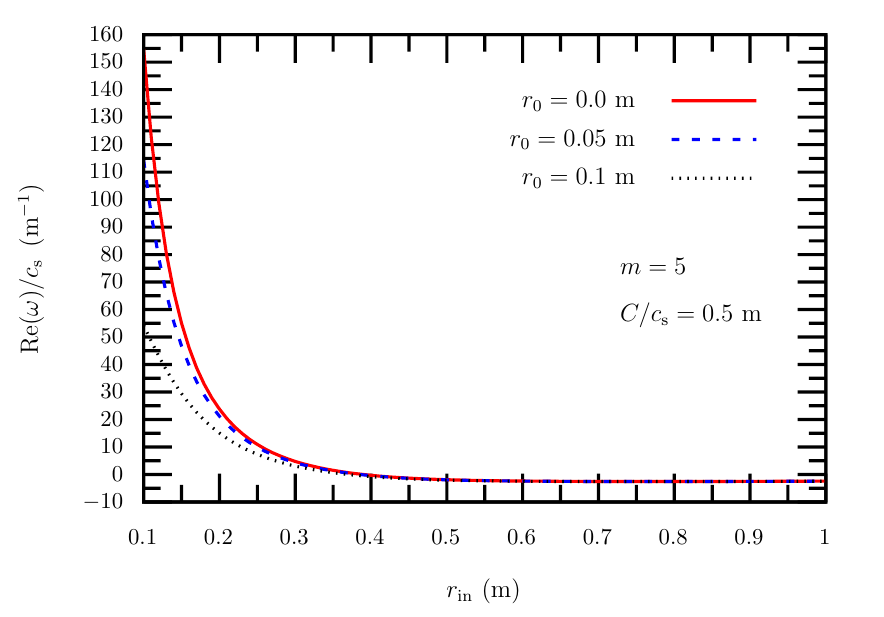}
\label{real_freq4}}\subfigure[Imaginary part of the fundamental~($n = 0$) QNM frequencies~$\omega$]{%
\includegraphics[width=0.5\textwidth]{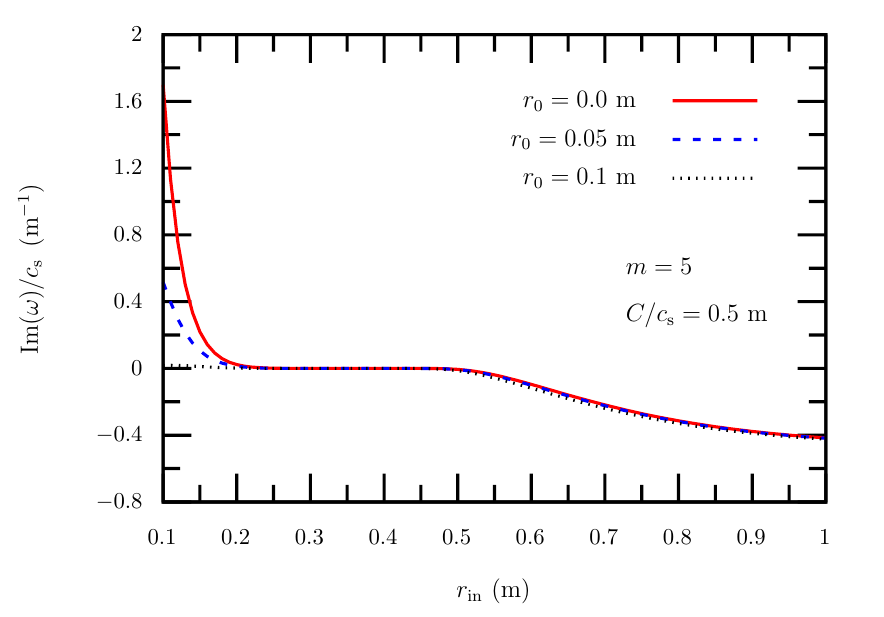}
\label{imag_freq4}}
\caption{Real and imaginary parts of the fundamental~($n = 0$) QNM frequencies~$\omega$, as functions of~$r_{\rm in}$, for azimuthal number~$m=5$ and circulation parameter~$C/c_{\rm s}=0.5\,\,{\rm m}$, for values of the vortex core~$r_0=0.0 \,\,{\rm m},\,0.05 \,\,{\rm m},\,{\rm and}\,\,0.1 \,\,{\rm m}$, obtained via continued-fraction method.}
\label{Freqs3}
\end{figure*}

In Fig.~\ref{Freqs3} we plot the real (left plots) and imaginary (right plots) parts of the fundamental~($n = 0$) QNM frequencies~$\omega$, as functions of~$r_{\rm in}$, for azimuthal number~$m=5$ and circulation parameter~$C/c_{\rm s}=0.5\,\,{\rm m}$, for different values of the vortex core~$r_0$, obtained via continued-fraction method. Here, we use the international system of units. Note that the vortex core~$r_0=0.0\,\,{\rm m}$ corresponds to the irrotational fluid flow. We can see that as we increase~$r_0$,~$\text{Im}(\omega)$ gets smaller. This suggests that the vorticity has the effect of quenching the instability. 

\begin{figure}[htpb!]
\centering
\includegraphics[width=0.5\textwidth]{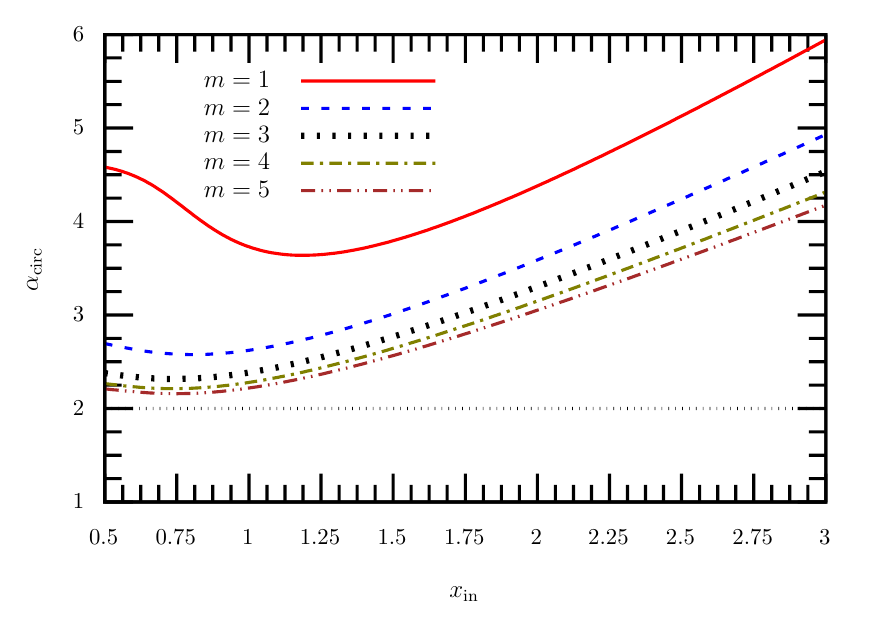}
\caption{Dimensionless circulation parameter~$\alpha_{\rm circ}$, as a function of~$x_{\rm in}$, corresponding to the configurations where the fundamental~($n = 0$) QNM frequencies~$\varpi$ go to zero, for azimuthal numbers $m=1,\,2,\,3,\,4,\,5$, obtained via continued-fraction method.}
\label{Zero_Freq}
\end{figure}

In Fig.~\ref{Zero_Freq} we plot the values of the dimensionless circulation parameter~$\alpha_{\rm circ}$, as a function of~$x_{\rm in}$, corresponding to the configurations where the fundamental~($n = 0$) QNM frequencies~$\varpi$ go to zero, for azimuthal numbers~$m=2,\,3,\,4,\,5,$ obtained via continued-fraction method. As we discussed before (cf.~Fig.~\ref{Freqs2}), as the azimuthal number~$m$ is increased the threshold between stability and instability increases. This may be seen  in Fig.~\ref{Zero_Freq} when the existence line of modes with~$\varpi=0$ approaches the smallest value of the dimensionless circulation parameter~$\alpha_{\rm circ}$ [cf. Eq.~\eqref{ergo_radius}], as the azimuthal number~$m$ is increased.

\section{Conclusions} 
\label{sec-Conclusion}
We have computed the QNMs of a rotating and incompressible fluid flow, considering the vorticity, using two different methods: direct integration and continued-fraction method. The system under consideration presents an analogue ergoregion but no analogue event horizon. This implies that, when we consider an inner boundary condition within the ergoregion, the system presents an instability, known as the ergoregion instability. We analyzed this instability for different values of the circulation of the fluid and of the position of the inner boundary.

We compared our results with the ones for the case without vorticity, finding that as we increase the vortex radius, the imaginary part of the frequency gets smaller. This indicates that the vorticity tends to diminish the instability of the system.

The ergoregion instability is a large-$m$ phenomenon, what is highlighted by the increased threshold between stability and instability as we increase the azimuthal number. Indeed, in this limit, the existence line for~$\varpi=0$, which separates the stable and unstable modes, approaches the~$\alpha_{\text{circ}}=2$ line, what corresponds to the minimum value for the existence of an ergoregion.

As examples of possible experimental implementations, we consider the sound waves in water (with speed~$c_{\rm s}=1493 \,\,{\rm m}/{\rm s}$) and waves propagating on the surface of water (with speed~$c_{\rm s}=\sqrt{g\,h_0}=0.78 \,\,{\rm m}/{\rm s}$)~(cf.~\cite{Torres:2016iee}). We estimate the time-scale of the ergoregion instabilities $t_{\rm scale}\equiv1/{\rm Im} \left(\omega\right)$ for azimuthal number~$m=5$, dimensionless circulation parameter~$\alpha_{\rm circ} =4.0$, imposing the boundary condition at~$x_{\rm in}=1.0$, obtained via continued-fraction method, considering an experimental value of the vortex core~$r_0=0.0134\,\,{\rm m}$~(cf.~\cite{Torres:2016iee}), namely
\beqn
&&t_{\rm scale}=0.0254\,\,{\rm s}\quad({\rm for\,\, sound\,\, waves}),\nn\\
\nn\\
&&t_{\rm scale}=48.4653\,\,{\rm s}\quad({\rm for\,\, surface\,\, waves}),\nn
\eeqn
which denotes a larger time-scale~$t_{\rm scale}$ of the ergoregion instabilities for surface waves when compared with sound waves.

Vorticity seems to play an important role in present and future experimental realizations of analogue models involving fluids. Therefore, studies of how it can affect such systems are required. In this work, we took a step in this direction, investigating the QNMs in a vortex with vorticity. A possible next step would be to consider not only the vorticity of the background, but also of the perturbation.

\section*{Acknowledgments}
The authors would like to acknowledge 
Funda\c{c}\~ao Amaz\^onia de Amparo a Estudos e Pesquisas (FAPESPA), 
Conselho Nacional de Desenvolvimento Cient\'ifico e Tecnol\'ogico (CNPq)
 and Coordena\c{c}\~ao de Aperfei\c{c}oamento de Pessoal de N\'ivel Superior (CAPES) -- Finance Code 001, from Brazil, for partial financial support.  
 This research has further been supported by the European Union's Horizon 2020 research and innovation (RISE) programme H2020-MSCA-RISE-2017 Grant No. FunFiCO-777740 and by the European Horizon Europe staff exchange (SE) programme HORIZON-MSCA-2021-SE-01 Grant No. NewFunFiCO-101086251.    
%


\end{document}